\documentclass[%
longbibliography,
amsmath,amssymb,
aps,
pra,
twocolumn,
floatfix,
]{revtex4-1}
\usepackage{graphicx}
\usepackage{dcolumn}
\usepackage{bm}
\usepackage{epsfig}
\usepackage{afterpage}

\usepackage[utf8]{inputenc}

\begin{document}

\title{Temperature Dependence of  the Density and Excitations of Dipolar Droplets}

\author{S. F. \"{O}zt\"{u}rk}
\email{sukrufurkanozturk@g.harvard.edu}
\affiliation{Department of Physics, Harvard University, Cambridge, Massachusetts, 02138, USA}
\author{Enes Aybar}
\altaffiliation[Present address: ]{ICFO-Institut de Ciencies Fotoniques, The Barcelona Institute of Science and Technology, 08860 Castelldefels (Barcelona), Spain }

\affiliation{Department of Physics, Bilkent University, Ankara 06800, Turkey}

\author{M. \"{O}. Oktel}
\affiliation{Department of Physics, Bilkent University, Ankara 06800, Turkey}

\date{\today}

\begin{abstract}
Droplet states of ultracold gases which are stabilized  by fluctuations have recently been observed for dipolar and two component Bose gases. These systems present a novel form of equilibrium where an instability at the mean field level is arrested by higher order correlations making the droplet states sensitive probes of fluctuations. In a recent paper, we argued that thermal fluctuations can play an important role for droplets even at low temperatures where the non-condensed density is much smaller than the condensate density. We used the Hartree-Fock-Bogoliubov theory together with local density approximation for fluctuations to obtain a generalized Gross Pitaevskii (GP) equation and solved it with a Gaussian variational ansatz to show that the transition between the low density and droplet states can be significantly modified by the temperature. In this paper, we first solve the same GP equation numerically with a time splitting spectral method to check the validity of the Gaussian variational ansatz. Our numerical results are in good agreement with the Gaussian ansatz for a large parameter regime and show that the density of the gas is most strongly modified by temperature near the abrupt transition between a pancake shaped cloud and the droplet. For cigar shaped condensates, as in the recent Er experiments, the dependence of the density on temperature remains quite small throughout the smooth transition. We then consider the effect of temperature on the collective oscillation frequencies of the droplet using both a time dependent Gaussian variational ansatz and real time numerical evolution. We find that the oscillation frequencies depend significantly on the temperature close to the transition for the experimentally relevant temperature regime ($\simeq 100$nK).

\end{abstract}
\maketitle

\section{Introduction}
\label{sec:introduction}
Ultracold bosonic gases form almost ideal, dilute, weakly interacting quantum systems. At temperatures much lower than the condensation temperature nearly all atoms share the collective ground state wavefunction, and the effect of interactions can be accounted for by their self-consistent action on that wavefunction. This mean-field idea, captured by the Gross-Pitaevskii equation \cite{pit61,gro61}, qualitatively describes all the equilibrium properties of almost all the trapped Bose gas experiments. The corrections to the dynamics described by GP equation arise from particles which are above the condensate. The effect of this condensate depletion is detectable, but remains small at low temperatures as long as the system remains dilute and weakly interacting. 

An alternative scenario for equilibrium, in which the effect of fluctuations is crucial, can arise if the mean field interactions describe an unstable system. The instability in the mean field description can be cured by the inclusion of fluctuations, which remain small compared to mean field effects even when they provide stability. This scenario of fluctuation stabilized equilibrium was first suggested theoretically by Petrov \cite{pet15} for two component Bose mixtures, and recently observed experimentally \cite{cts18,sfm18,ebp19}. For dipolar gases, droplet formation in the mean-field unstable regime was first reported experimentally, where follow-up experiments have definitively established the fluctuation based stabilization as their formation mechanism \cite{ksw16,fks16,swb16,cbp16}.   

The theoretical description of the dipolar droplets have so far \cite{wsa16-1,wsa16-2,bbl15,bwb16-1,bwb16-2} relied on the generalized Gross Pitaevskii equation for the condensate wave function
\begin{multline}
\left[ - \frac{\hbar^2\nabla^2}{2 m} + U_{tr}(\mathbf{x}) + \int d^{3}\textbf{x}'V_{\text{int}}(\textbf{x},\textbf{x}') 
  |\Psi(\textbf{x}')|^2 \right] \Psi(\mathbf{x}) \\ = \left(\mu-\Delta\mu_{QF}(\mathbf{x})\right) \Psi(\mathbf{x}) ,
\end{multline}
where the usual GP equation is modified by the quantum fluctuation contribution to chemical potential $\Delta\mu_{QF}(\mathbf{x})$.
This contribution is then assumed to be determined entirely in terms of the local condensate wavefunction as it would be found in a uniform gas
\begin{equation}
\Delta\mu_{QF}(\Psi(\mathbf{x})) \propto |\Psi(\mathbf{x})|^{3}.
\end{equation}
The exact coefficient linking $|\Psi|^3$ to $\Delta\mu_{QF}$ depends on the infrared cutoff choice used the in the local density approximation, but all different choices used in the literature  for this cutoff give results which are in agreement with observations within experimental accuracy. 

In a recent paper \cite{aok19}, we showed that the modified GP equation can be derived using the Hartree-Fock-Bogoliubov theory (HFBT) \cite{gri96} together with local density approximation for fluctuations. Such a treatment makes the assumptions underlying the generalized GP equation more transparent. We found that  the above form of the generalized GP equation is valid only when the Hartree potential created by the depleted particles is negligible. This condition is satisfied in the dipolar droplet experiments, where the estimates for the number of depleted particles remain small even in the droplet regime.

The novel physical regime probed in the droplet experiments is evident in this set of approximations, as the number of depleted particles is negligible compared to the number of condensed particles, but their interaction with the condensate wavefunction supply the energy cost required for stability.

The derivation of the generalized GP equation from HFBT can be carried out at finite temperature \cite{aok19,bou17} without further complications. Quantum fluctuations are supplemented by thermal fluctuations, and in general the dynamics of such a system should be given in terms of a two fluid model with separate equations for the condensate and normal components. However, at low enough temperatures, the total number of depleted atoms would still be much smaller than the number of atoms in the condensate, and the dynamics of the fluctuations can be described only in terms of the local condensate wavefunction. Thus, a finite temperature version of the generalized GP equation was derived in which local chemical potential correction is now a function of both the condensate density and the temperature $\Delta\mu_{QF}(\Psi(\mathbf{x}),T)$. We solved this equation with a Gaussian variational ansatz and found that even if the temperature depletion remains small, the condensate density may be strongly modified. The change in the local compressibility is particularly pronounced near the first order transition to  the droplet state. 
 
In this paper, we first verify the conclusions obtained in Ref.\cite{aok19}, with a numerical solution of the generalized GP equation. We investigate the effects of temperatures which are much smaller than the BEC transition temperature, so that the thermal depletion density is much smaller than the condensate density but comparable to the quantum depletion. This temperature regime is likely to be relevant for both the Dy experiments by the Stuttgart group \cite{ksw16,fks16,swb16}, and the Er experiments by the Innsbruck group \cite{cbp16}. Our numerical results show that the effect of temperature on the density is most prominent for sudden transitions to the droplet states as obtained for initially oblate (pancake) clouds. For cigar shaped (prolate) clouds, the transition to the droplet state is smooth, and the effect of temperature remains small throughout the transition. Based on this observation, we investigate whether collective oscillation frequencies show any dependence on temperature at these low temperatures. We use both an effective Lagrangian based on a Gaussian ansatz, and real time numerical evolution to investigate the lowest lying collective excitations of the droplet state. We find that the collective oscillation frequencies show strong dependence on temperature close to the first order transition. 


In the next section, we briefly summarize our HFBT method and give the resulting temperature dependent generalized GP equation. Section \ref{sec:numerical} explains our numerical algorithm and gives the calculated equilibrium density profiles for the Er and Dy experiments. Section \ref{sec:oscillation} contains the temperature dependence of the lowest collective oscillation frequencies of the condensate obtained by both a variational Lagrangian method and numerical real time evolution. A summary of the results and their relevance for the experiments on dipolar droplets are given in section \ref{sec:experiments}.

\section{Hartree Fock Bogoliubov theory and the modified Gross Pitaevskii Equation}
\label{sec:HFBT}
We consider Bosonic atoms of mass $M$ in a confining potential $U_{\text{tr}}(\mathbf{x})$ with chemical potential $\mu$,  described by the Hamiltonian
\begin{eqnarray}\label{Hamiltonian} \hat{H}&=&\int d^{3} \mathbf{x}
\hat{\psi}^{\dagger}(\mathbf{x}) \left( -\frac{\hbar^{2}\nabla^{2}}{2M}+U_{\text{tr}}(\mathbf{x})-\mu\right) \hat{\psi}(\mathbf{x})\\ & +
&\frac{1}{2} \iint d^{3} \mathbf{x} d^{3} \mathbf{x}'
\hat{\psi}^{\dagger}(\mathbf{x})
\hat{\psi}^{\dagger}(\mathbf{x}')V_{\text{int}}(\mathbf{x}-\mathbf{x}')
\hat{\psi}(\mathbf{x}') \hat{\psi}(\mathbf{x}), \nonumber
\end{eqnarray}
with the Bosonic commutation $[\hat{\psi}(\mathbf{x}),\hat{\psi}^{\dagger}(\mathbf{x}')]=\delta(\mathbf{x}-\mathbf{x}')$ of field operators. The interaction potential between the particles have both a short range, and a long range (dipolar) part
\begin{equation}
 V_{\text{int}}(\mathbf{x})=g\left[\delta(\mathbf{x}) +\frac{3\epsilon_{dd}}{4\pi |\mathbf{x}|^{3}}\left(1-3\frac{z^2}{|\mathbf{x}|^{2}}\right) \right],
\end{equation}
where the short range repulsion is expressed in terms of the s-wave scattering length $a_s$ as $g=4\pi\hbar^{2}a_{s}/M$ and the relative strength of the dipolar interaction defines the dimensionless parameter $ \epsilon_{dd}$. 

We consider systems where most of the atoms occupy the condensate state. The total particle number $N$  is set by the chemical potential and $N_0$, the largest eigenvalue of the one particle density matrix $\langle \hat{\psi}^\dagger \hat{\psi} \rangle$ is close to this number $N-N_{0}\ll N$. The part of the field operator corresponding to the condensate can then be safely approximated by the classical field $ \Psi(\textbf{x})$, the condensate wavefunction. The remaining part, which is referred to as the fluctuation operator, $ \hat{\phi}(\textbf{x})=\hat{\psi}(\textbf{x})-\Psi(\textbf{x})$ can be used to define the one particle non-condensate density matrices. The direct non-condensate density is
\begin{equation}
\tilde{n}(\textbf{x}',\textbf{x})=\langle\hat{\phi}^{\dagger}(\textbf{x}')\hat{\phi}(\textbf{x})\rangle,
\end{equation}
and the anomalous non-condensate density is
\begin{equation}
\tilde{m}(\textbf{x}',\textbf{x})=\langle\hat{\phi}(\textbf{x}')\hat{\phi}(\textbf{x})\rangle.
\end{equation}
\begin{widetext}
The aim of HFBT is to reduce the Hamiltonian to a system of self-consistent equations which describe the equilibrium values for the condensate wavefunction $\Psi(\textbf{x})$ and the non-condensate densities $\tilde{n},\tilde{m}$ \cite{gri96}. This is achieved by assuming that the higher order correlation functions can be factorized in terms of these three quantities, e.g. 
\begin{equation}
\hat{\phi}^{\dagger}(\textbf{x})\hat{\phi}^{\dagger}(\textbf{x}')
  \hat{\phi}(\textbf{x}') \approx
 \tilde{m}^{*}(\textbf{x}',\textbf{x})\hat{\phi}(\textbf{x}') +
\tilde{n}^{*}(\textbf{x}',\textbf{x})\hat{\phi}^{\dagger}(\textbf{x}') +
   \tilde{n}(\textbf{x}')\hat{\phi}^{\dagger}(\textbf{x}). 
\end{equation}
This procedure yields the GP equation
\begin{equation}
  \mathcal{L}\Psi(\textbf{x})+\int d^{3}\textbf{x}'V_{\text{int}}(\textbf{x}-\textbf{x}')
  \tilde{n}(\textbf{x}',\textbf{x})\Psi(\textbf{x}')+
  \int d^{3}\textbf{x}'V_{\text{int}}(\textbf{x}-\textbf{x}')\tilde{m}(\textbf{x}',\textbf{x})\Psi^{*}(\textbf{x}')=0,
 \end{equation}
 where
$  \mathcal{L}=\left[-\hbar^{2}\nabla^{2}/2M-\mu+U_{\text{tr}}(\textbf{x})+\int d^{3}\textbf{x}'V_{\text{int}}(\textbf{x}-\textbf{x}')|\Psi(\textbf{x}')|^{2}
  +\int d^{3}\textbf{x}'V_{\text{int}}(\textbf{x}-\textbf{x}')\tilde{n}(\textbf{x}')\right],$ displaying the dependence of the condensate wavefunction on the non-condensate densities.
	
To complete the self consistent loop, non-condensate densities must be calculated in terms of the condensate wavefunction. This is achieved by 
obtaining the Bogoliubov-DeGennes (BdG) equations for excitations,
\begin{equation}\label{eq:BdGu}
\begin{split}
  \mathcal{L}_{0} u_{j}(\textbf{x})+\int d^{3}\textbf{x}'V_{\text{int}}(\textbf{x}-\textbf{x}')
  \Psi^{*}(\textbf{x}')\Psi(\textbf{x}) u_{j}(\textbf{x}')
 -\int d^{3}\textbf{x}'V_{\text{int}}(\textbf{x}-\textbf{x}')
\Psi(\textbf{x}')\Psi(\textbf{x}) v_{j}(\textbf{x}')
 =E_{j}u_{j}(\textbf{x}) \\ 
  \mathcal{L}_{0} v_{j}(\textbf{x})+\int d^{3}\textbf{x}'V_{\text{int}}(\textbf{x}-\textbf{x}')
  \Psi(\textbf{x}')\Psi^{*}(\textbf{x}) v_{j}(\textbf{x}')
 -\int d^{3}\textbf{x}'V_{\text{int}}(\textbf{x}-\textbf{x}')
\Psi^{*}(\textbf{x}')\Psi^{*}(\textbf{x}) u_{j}(\textbf{x}')
 =-E_{j}v_{j}(\textbf{x}),
 \end{split}
 \end{equation}
 where $  \mathcal{L}_{0}=\left[-\hbar^{2}\nabla^{2}/2M-\mu+U_{\text{tr}}(\textbf{x})+\int d^{3}\textbf{x}'V_{\text{int}}(\textbf{x}-\textbf{x}')|\Psi(\textbf{x}')|^{2}\right].$  The solutions for the amplitudes $u,v$ can be summed to give the non-condensate densities
 \begin{equation}\label{eq:ntildeuv} \begin{split}
   \tilde{n}(\textbf{x}',\textbf{x})=&\sum_{j}\left(v_{j}(\textbf{x}')v_{j}^{*}(\textbf{x})
   +N_{B}(E_{j})\left[u_{j}^{*}(\textbf{x}')u_{j}(\textbf{x})+v_{j}(\textbf{x}')v_{j}^{*}(\textbf{x})\right]\right)\\
   \tilde{m}(\textbf{x}',\textbf{x})=&-\sum_{j}\left(u_{j}(\textbf{x}')v_{j}^{*}(\textbf{x})
   +N_{B}(E_{j})\left[v_{j}^{*}(\textbf{x}')u_{j}(\textbf{x})+u_{j}(\textbf{x}')v_{j}^{*}(\textbf{x})\right]\right).
   \end{split}
 \end{equation}
It is important to note here that we have ignored the interactions between the non-condensate particles in the above Eq.\ref{eq:BdGu}, as we are interested at temperatures much lower than condensation temperature and the depletion remains small.

The calculation of non-condensate densities in terms of the condensate wavefunction in principle requires the determination of the full spectrum of the BdG equations. While the first few modes of these equations would have wavelengths close to the size of the system, the wavelength of the following BdG modes would get much smaller than the size of the system, or the coherence length of the condensate. For such modes the condensate density and the external trapping potential are slowly varying and  Eq.\ref{eq:BdGu} can be solved with a local density approximation akin to the WKB approximation. 
Within this approximation, the non-condensate densities are locally determined in terms of the condensate wavefunction. The self consistency requirement can be stated only in terms of the condensate wavefunction, which yields the generalized GP equation.

At zero temperature, the condensate wavefunction then satisfies
 \begin{equation}\label{mGP} \left[\left( -\frac{\hbar^{2}\nabla^{2}}{2M}+U_{\text{tr}}(\mathbf{x})-\mu\right)+\Phi_{H}(\mathbf{x})
+\Omega_{n}(\mathbf{x}) +\Omega_{m}(\mathbf{x}) \right] \Psi(\mathbf{x})=0,
\end{equation}
where the Hartree potential is
$\Phi_{H}(\mathbf{x})=\int d^{3}\mathbf{x}'
V_{\text{int}}(\mathbf{x}-\mathbf{x}')(|\Psi(\mathbf{x}')|^{2}+\tilde{n}(\mathbf{x}'))$. The local fluctuation induced potential, which is equivalent to the Lee-Huang-Yang correction has two contributions coming from direct and anomalous non-condensate densities, 
\begin{eqnarray}\label{eq:QFterms}
\Omega_{n}(\textbf{x})\Psi(\textbf{x}) &=& \frac{8}{3}gn_{0}(\textbf{x})\sqrt{\frac{a^{3}_{s}n_{0}(\textbf{x})}{\pi}}
\mathcal{Q}_{5}(\epsilon_{dd})\Psi(\textbf{x}), \\
\Omega_{m}(\textbf{x})\Psi(\textbf{x}) &=&8gn_{0}(\textbf{x})\sqrt{\frac{a^{3}_{s}n_{0}(\textbf{x})}{\pi}}
\mathcal{Q}_{5}(\epsilon_{dd})\Psi(\textbf{x}), \nonumber \end{eqnarray} where $
Q_{l}(\epsilon_{dd})=\int_{0}^{1} du [1+\epsilon_{dd}(3u^{2}-1)]^{l/2} $.
Explicitly, this is almost the generalized GP equation used in the literature \cite{wsa16-1,bwb16-1}, 
\begin{equation}\label{explicitGP}
\left[-\frac{\hbar^{2}\nabla^{2}}{2M}+U_{\text{tr}}(\mathbf{x})-\mu+\int
d^{3}\mathbf{x}'V_{\text{int}}(\mathbf{x}-\mathbf{x}')\left(|\Psi(\mathbf{x}')|^{2}+\tilde{n}(\mathbf{x}')\right)+\frac{32}{3}g\sqrt{\frac{a^{3}_{s}}{\pi}}
\mathcal{Q}_{5}(\epsilon_{dd})|\Psi(\mathbf{x})|^{3}\right]\Psi(\mathbf{x})=0.
\end{equation}
Thus the only correction to the generalized GP equation obtained by adding local LHY correction to the chemical potential is the Hartree potential created by the depleted particles. If the depletion $\tilde{n}(\mathbf{x})=\frac{8}{3}\sqrt{\frac{a^{3}_{s}}{\pi}}
\mathcal{Q}_{3}(\epsilon_{dd})|\Psi(\mathbf{x}')|^{3}$ remains small, this contribution will be accordingly small.  
\end{widetext}

If the dipolar interaction is not dominant, $\epsilon_{dd}<1$, the trapped gas will be stable at low densities and fluctuation contribution is not qualitatively important. However for strong dipolar interaction, $\epsilon_{dd}>1$, the gas has an instability towards collapse as the attractive part of the dipolar interaction can favor higher densities. The local fluctuation induced interaction potential can arrest this collapse as it scales with a higher power of the density, $n_0^{3/2}$, compared to the direct interaction with $n_0$. While this general physical mechanism has been tested experimentally \cite{fks16} to be at play in the droplet experiments, the exact coefficient of the fluctuation term is not easy to determine. When $\epsilon_{dd}>1$ there is no stable solution for a uniform system, some long wavelength modes will have imaginary frequencies signaling collapse. Thus the expressions $\Omega_n$, and $\Omega_m$ (Eq. \ref{eq:QFterms}) will have imaginary parts. One approach in the literature was to ignore the imaginary components, assuming they only cause decay at long times \cite{swb16}. However this imaginary part is clearly a by--product of the LDA, which as we have discussed above is valid only for higher Bogoliubov modes. A careful accounting of the discrete nature of the Bogoliubov modes would not show any imaginary frequencies, but a computationally much simpler approach is to impose a cutoff on which LDA modes should be taken into account as employed in Ref\cite{wsa16-1,bwb16-1}. We used a spherical cutoff in k-space which excludes all modes with wavevectors smaller than $k_{c}=\frac{\pi}{2\xi}$, as the coherence length of the condensate $\xi$ is the length scale which characterizes locality for the condensate. 

The systematic derivation of the generalized GP equation using HFBT enables straightforward generalization to finite temperatures. While the HFBT is mostly reliable at all temperatures, local calculation of non-condensate densities solely in terms of local condensate wavefunction puts a severe restriction on the applicability of LDA to obtain a generalized GP equation at non-zero temperature. The dynamics of condensate and non-condensate atoms are controlled by different equations, and only when the number of non-condensate atoms is small enough their dynamics will be completely determined by the condensate. At zero temperature this essentially means that the system is weakly interacting enough to limit quantum depletion. At finite temperature non-condensate density is contributed to by both quantum and thermal depletion mechanisms. The non-condensate part dynamics will be controlled completely by the condensate only if the total depletion remains small. Essentially, to be able to write a single self consistent equation  for the condensate the temperature must be much smaller than the BEC transition temperature so that the thermal depletion is comparable to the quantum depletion.   
\begin{widetext}

At such low but non-zero temperatures, the generalized GP equation takes the similar form
\begin{equation} \label{eq:GPT}
\left[ - \frac{\hbar^2\nabla^2}{2 m} + U_{tr}(\mathbf{x}) + \int d^{3}\textbf{x}' V_{\text{int}}(\textbf{x}-\textbf{x}')
  |\Psi(\textbf{x}')|^2 \right] \Psi(\mathbf{x})+ \Delta\mu_{QF}(\mathbf{x},T) \Psi(\mathbf{x}) = \mu \Psi(\mathbf{x}),
\end{equation}
with temperature only playing a role in the determination of the local fluctuation induced interaction. With the same cutoff used at zero temperature we obtain
\begin{equation}
\Delta\mu_{QF}(\mathbf{x},T) = \frac{32}{3}g\sqrt{\frac{a_s^{3}}{\pi}}\left(\mathcal{Q}_{5}(\epsilon_{dd})
+\mathcal{R}(\epsilon_{dd},t(\mathbf{x}))\right)|\Psi(\textbf{x})|^{3},
\end{equation}
with dimensionless temperature $t=k_B T/g n_0(\mathbf{x})$ and 
\begin{equation} \mathcal{Q}_{5}(\epsilon_{dd};q_{c})=\frac{1}{4\sqrt{2}}\int_{0}^{1}du f(u)\left[\left(4f(u)-q_{c}^{2}\right)\sqrt{2f(u)+q_{c}^{2}}-3f(u)q_{c}+q_{c}^{3}\right] \end{equation}
\begin{equation} \mathcal{R}(\epsilon_{dd},t;q_{c})=\frac{3}{4\sqrt{2}}\int_{0}^{1}du\int_{q_{c}^{2}}^{\infty}dQ\frac{Qf(u)}{\sqrt{Q+2f(u)}}\frac{1}{\exp[\sqrt{Q\left(Q+2f(u)\right)}/t]-1}. \end{equation}

In Ref. \cite{aok19} we calculated these dimensionless functions, and obtained a $t^2$ power law fit  
$\mathcal{R}(\epsilon_{dd},t)=S(\epsilon_{dd})t^{2}$ and the coefficient is best approximated by  
$\mathcal{S}(\epsilon_{dd})=-0.01029\epsilon_{dd}^{4}+0.02963\epsilon_{dd}^{3}-
0.05422\epsilon_{dd}^{2}+0.009302\epsilon_{dd}+0.1698$ for $0<\epsilon_{dd}<2$.
\end{widetext}

\section{Numerical Calculation of the Density profiles}
\label{sec:numerical}

We now discuss the numerical solution of Eq.\ref{eq:GPT} with the parametrization in the previous section, and compare it with the variational solution presented in Ref.\cite{aok19}. Our focus is on the observable effects of temperature on the density profile. As our LDA approach neglects the effect of interactions of the non-condensate atoms the temperature range will be limited so as to make sure that the condensate fraction remains close to one.

We use time splitting spectral method \cite{bjm03} to evolve the GP equation in imaginary time until convergence in energy is reached. A three dimensional grid of up to 64 by 64 by 256 points in real space is used to capture the anisotropic shape of the droplets, where the real space discretization step is determined to ensure smooth variation of the cloud density.

\begin{figure}
\centering

    \includegraphics[scale=0.55]{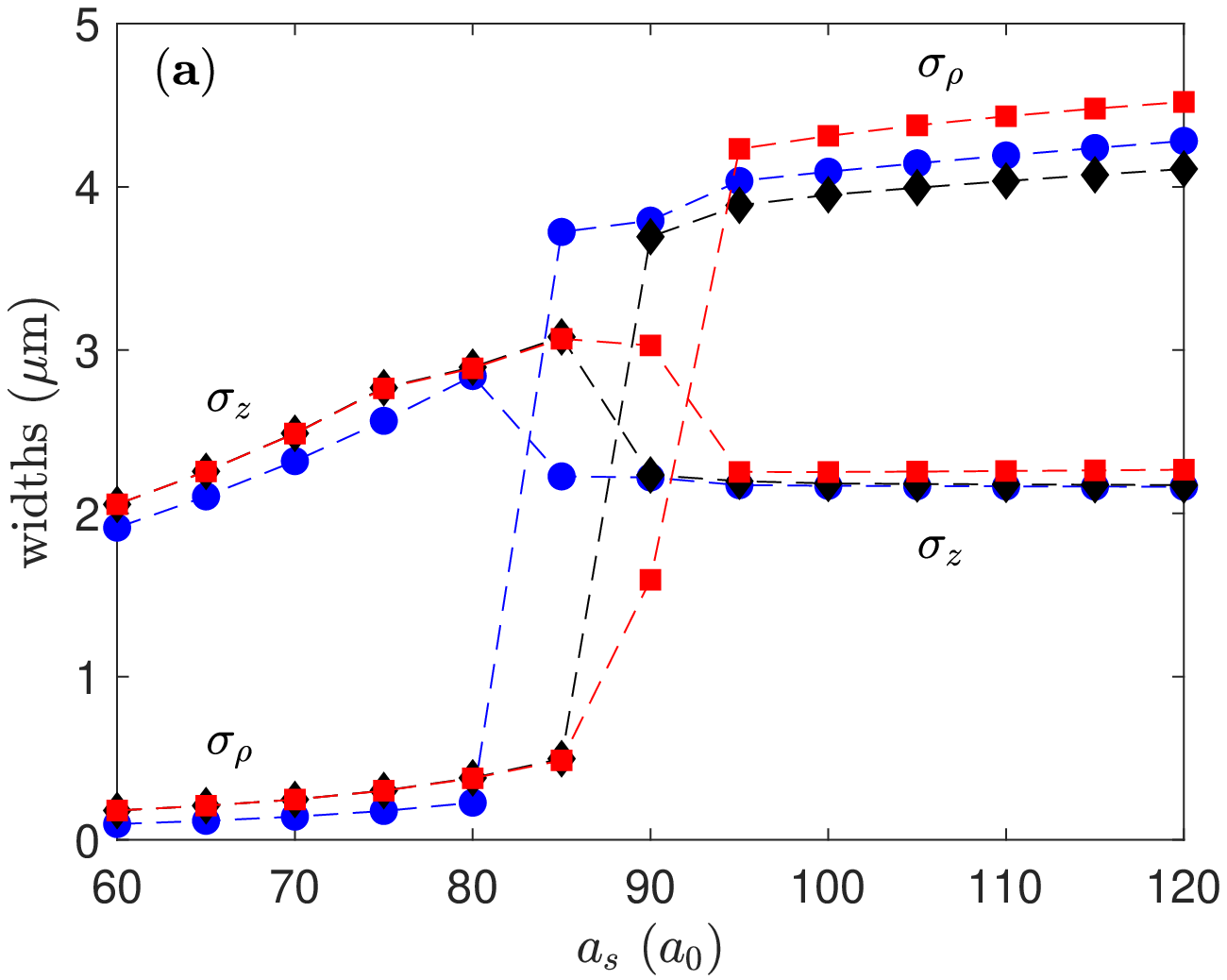}
    \includegraphics[scale=0.55]{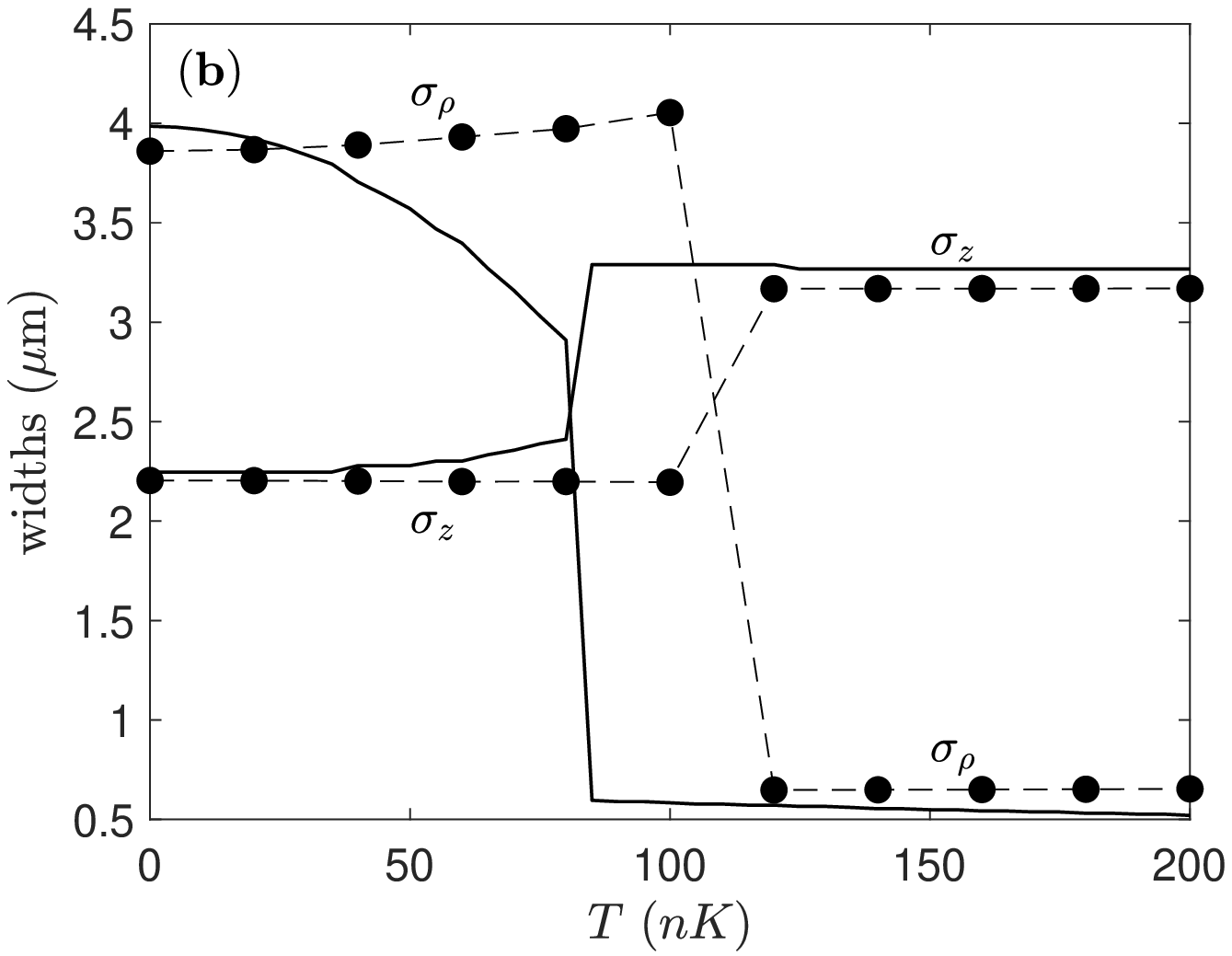}
    \caption{(a) Radial and axial widths, $\sigma_{\rho}$ and $\sigma_{z}$, of a Dy BEC with $N=2000$ atoms in a pancake shaped trap with harmonic frequencies $\left\{\omega_\rho, \omega_z\right\}=2\pi\times\left\{45,133\right\}$ Hz as a function of the s-wave scattering length $a_s$. The dots, diamonds and squares correspond to the temperatures $T=0$, $T=70$nK, and $T=150$nK respectively. The dashed lines are guide for the eye and the points are obtained by numerical time evolution. The cloud density and the transition point to a high density droplet phase is significantly altered as the temperature is varied. At a critical $a_s$ value of $88a_0$ (b) shows the change of radial and axial widths as a function of the temperature. The dots are showing the results of numerical time evolution and the solid line is obtained by assuming a Gaussian ansatz.}
    
    \label{fig:DyWidths}
\end{figure}

\begin{figure*}
\centering

\includegraphics[width=0.8\textwidth]{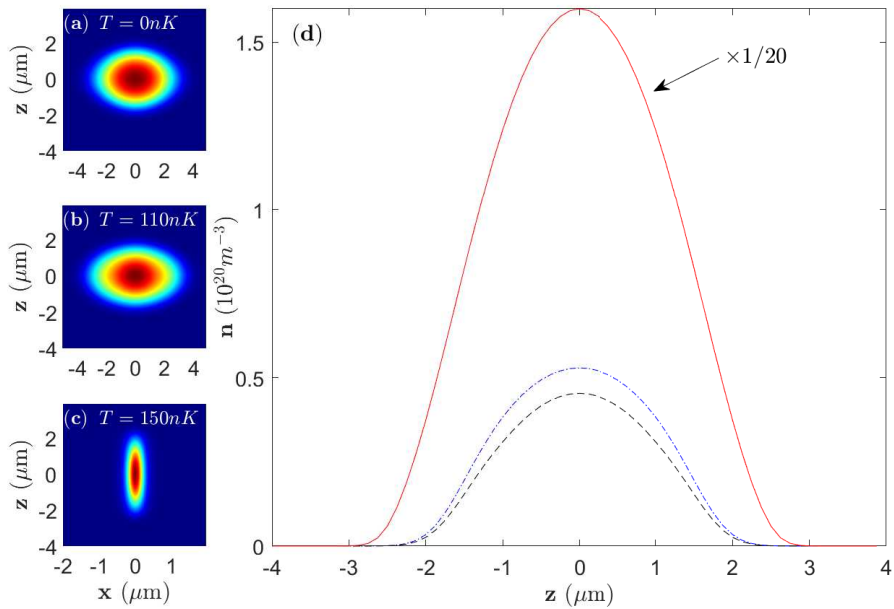}
\caption{(a),(b),(c) Density profile $(x, z, y=0)$ of a Dy BEC with $N=2000$ atoms and $a_s=88$ at $T=0$, $T=110$nK, $T=150$nK respectively, in a pancake shaped trap with harmonic frequencies $\left\{\omega_\rho, \omega_z\right\}=2\pi\times\left\{45,133\right\}$ Hz. The 2D profiles show the transition from a BEC to a high density droplet as the temperature is increased. (d) shows the cuts along the z direction with $x,y=0$. The blue (dash-dotted), black (dashed) and red (solid) lines correspond to the temperatures $T=0$, $T=110$nK, and $T=150$nK respectively. For visual clarity the density cut for the $T=150$nK case (red) is reduced by a factor of 20.}
\label{fig:GaussianDensity}
\end{figure*}

\begin{figure*}

\centering
\includegraphics[width=0.9\textwidth]{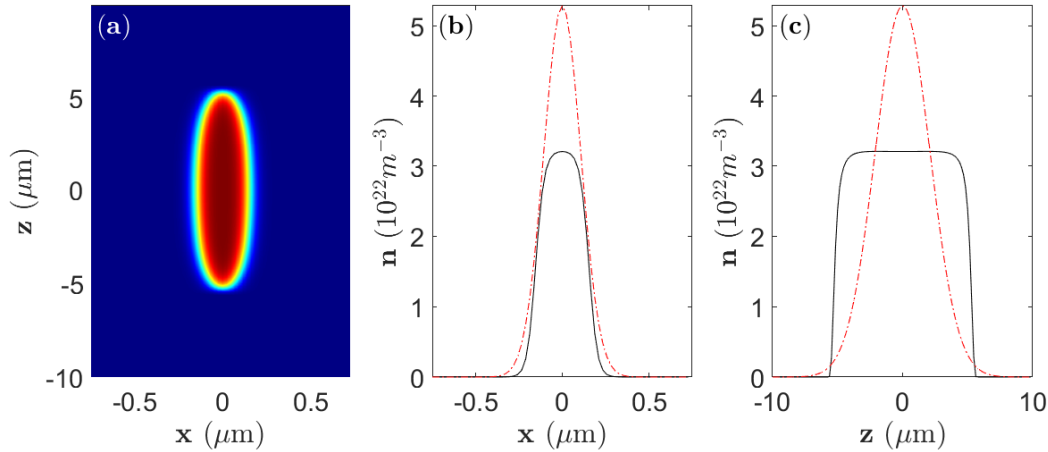}
\caption{(a) Density profile $(x, z, y=0)$ of a high density Dy droplet with $N=20000$ atoms and $a_s=60$ in a spherical trap with $\nu_i = 70$ Hz at $T=0$, obtained by imaginary time evolution. (b) and (c) show the cuts along the x and z direction with $y,z=0$ and $x, y=0$ respectively. The black (solid) lines are obtained by imaginary time evolution and the flat top profile of the high density droplets is apparent. The red (dash-dotted) lines show the analytical results assuming a Gaussian ansatz. The difference between the ground state density profiles manifests itself in the lowest-lying excitations. The axial oscillation frequencies for the exact numerical density profile and the Gaussian ansatz are $5.01 \nu_z$ and $3.96 \nu_z$ respectively, where $\nu_z$ is the axial trap frequency.}

\label{fig:FlatDropletDensity}
\end{figure*}
The Hartree potential created by the long range dipolar interaction is calculated by expressing it as a convolution in k space. 
\begin{equation}
V_{H}^{(dd)}(\mathbf{x}) = \int \frac{d^{3}\mathbf{k}}{\left(2\pi\right)^{3}}\tilde{n}_{0}(\mathbf{k})\tilde{V}_{dd}(\mathbf{k})e^{i\mathbf{k}\cdot\mathbf{x}},
\end{equation}
where the Fourier transform of the dipolar interaction is 
\begin{equation}
\tilde{V}_{dd}(\mathbf{k})=\epsilon_{dd}\left(3\frac{k_{z}^{2}}{|\mathbf{k}|^{2}}-1\right).
\end{equation}

While this expression avoids the complications of the dipolar potential near zero separation the use of discrete Fourier transforms instead of the continuum integrals effectively introduces repeated copies of the simulation cell. In order to avoid the interaction between neighboring real space cells we used a spherical cutoff on the dipolar interaction which sets the interaction to zero beyond the cutoff distance. This cutoff distance must be chosen larger than the maximum size of the droplet so that the points at the edges of the cigar shaped cloud interact with each other. However, it must be shorter than then the distance between the closest points of clouds in spuriously repeated simulation cells, so that long range interactions take place only within one simulation cell.  We choose the size of the simulation cell and the cutoff to comply with these constraints.   
The Fourier transform of the dipolar potential with this cutoff is calculated as 
\begin{equation}
\tilde{V}_{dd}^{(cut)}(\mathbf{k})=\tilde{V}_{dd}(\mathbf{k})\left(1-3\frac{\sin\left(|\mathbf{k}|R\right)-|\mathbf{k}|R\cos\left(|\mathbf{k}|R\right)}{\left(|\mathbf{k}|R\right)^{3}}\right)
\end{equation}

After numerically obtaining the solution, we extract the widths using the moments such that they correspond to the variational width parameters when applied to a variational Gaussian state \cite{bbl15}, 
\begin{eqnarray}
\sigma_{\rho}^2=\frac{4}{N}\int d^3\mathbf{x} \rho^2 |\Psi(\mathbf{x})|^2, \\ \nonumber
\sigma_{z}^2=\frac{2}{N}\int d^3\mathbf{x} z^2 |\Psi(\mathbf{x})|^2.
\end{eqnarray}

We first consider a parameter regime for which there is a sharp transition from a pancake shaped low density cloud to a single droplet. Focusing on the parameter regime explored in the Stuttgart experiments \cite{ksw16,fks16,swb16},  we calculate the density profile for $N=2000$ Dy atoms in an axially symmetric trap with frequencies $f_\perp=45$Hz, $f_z= 133$Hz. At zero temperature the cloud enters the droplet state below $a_c^{(0)}=83 a_0$. As can be seen in Fig.\ref{fig:DyWidths}, increasing temperature shifts this transition to higher scattering lengths, where a 100nK change in the cloud temperature can be observed to make a $\sim 10 a_0$ change in the transition point. It is possible to induce the transition from the cloud to the droplet by increasing the temperature as shown in Fig \ref{fig:DyWidths}, for $a_s=88 a_0$ the cloud enters the droplet state around $T \simeq 110nK$.

We see from our derivation that there are three effects of raising the temperature in a dipolar droplet system. First effect will be that the LHY term increases as thermal fluctuations now supplement quantum fluctuations stabilizing the droplet state. The second effect is that the number of particles in the condensate is lowered because of the thermal depletion, and following this the third effect is that the depleted particle density will have long range dipolar interactions with the condensate.  Our calculations show that at the lowest temperatures which are relevant for the current experiments the change in the compressibility of the droplet, i.e. the LHY term, will be the most dominant of these three effects.  However, even though the temperature contribution to LHY term is always positive,  its effect on the transition between the low density and the droplet state is non-trivial.  

Let’s particularly focus on the pancake shape traps for which the distinction between the droplet and low density phases is sharp.  The attractive part of the dipolar interaction favors higher density and higher aspect ratios for the cloud, which would normally collapse the cloud for $\epsilon_{DD}>1$. This inclination is resisted by two different mechanisms for the low density and droplet phases. In the low density phase the trap potential along the z direction combined with the short range repulsion stops the collapse. The effect of low temperature is negligible for the low density minimum.  For the droplet phase the confining potential is irrelevant and the collapse is arrested by the LHY energy which increases with a higher power of density. The first effect of temperature is an increase in the coefficient of the LHY term. 

In a pancake trap the transition is first order, so the system chooses the local minimum with the lower energy. As temperature is increased the minimum energy corresponding to the low density phase barely changes, while the minimum energy for the droplet changes due to the change in the LHY term. While the LHY contribution to energy increases, the droplet aspect ratio, thus the negative contribution of the dipolar interaction energy also increases. Increasing temperature lowers the total energy of the droplet  minimum, resulting in the transition to the droplet state seen in Fig.\ref{fig:DyWidths}.   While we base this argument on the variational study in \cite{aok19},  numerical results obtained here show that the non-trivial effect of the temperature on the transition is not an artifact of the restrictions to Gaussian wavefunctions. The results of the numerical calculation are in good agreement with the variational calculation presented in ref.\cite{aok19}. As can be seen Fig.\ref{fig:GaussianDensity} even though the density of the center changes by a factor of $\simeq 20$ the profile of the cloud remains well approximated by a Gaussian throughout the transition, thus the variational fits retain their predictive power.

We also simulate what happens in the deep droplet state as can be seen in Fig. \ref{fig:FlatDropletDensity}, the density profile is not a Gaussian any more, and the numerical solution is different from the variational result. The effect of low temperatures on this strong droplet state remains negligible.

\begin{figure}

\centering
    \includegraphics[width=0.50\textwidth]{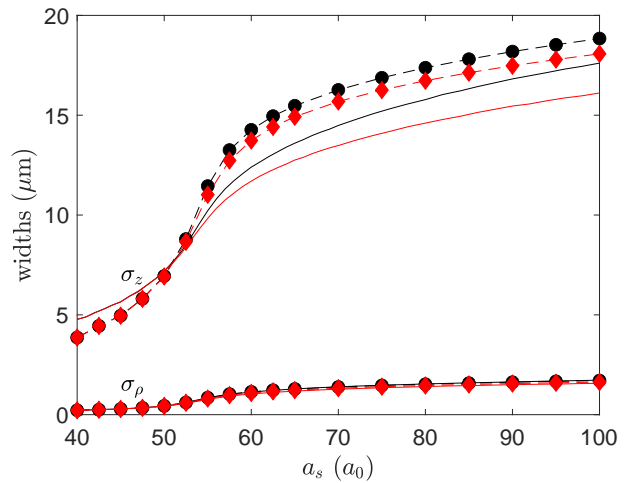}
    \caption{(a) Radial and axial widths, $\sigma_{\rho}$ and $\sigma_{z}$, of an Er BEC with $N=20000$ atoms in a cigar shaped trap with harmonic frequencies $\left\{\omega_\rho, \omega_z\right\}=2\pi\times\left\{178,17\right\}$ Hz as a function of the s-wave scattering length $a_s$. The black line and dots correspond to the temperature $T=0$, the red line and diamonds correspond to $T=150$nK. The dashed lines in the figure are guide for the eye. The solid lines are obtained analytically by assuming a Gaussian ansatz and the points are the results of the numerical time evolution. The finite temperature effects are observable yet not significant for the cigar shaped droplets.}
    
    \label{fig:ErWidths}
\end{figure}

Next, we investigate a parameter regime for which the transition between the droplet and low-density phases is smooth, as observed in the Innsbruck experiments Ref. \cite{cbp16}. For $20000$ Er atoms in a trap with frequencies $f_\perp=178$Hz, $f_z= 17.2$Hz the transition to the droplet is a continuous crossover near $a_s\simeq 50 a_0$. We find that the temperatures up to $200 nK$ have a distinct, yet small, effect on the cloud radii throughout, as can be seen in Fig. \ref{fig:ErWidths}. Once again the density profile of the cloud remains Gaussian throughout the transition and the variational results are in good agreement with the numerical calculations. 

\section{Oscillation Frequencies of droplets at finite temperature}
\label{sec:oscillation}
Collective oscillations of trapped condensates have emerged as one of the most valuable experimental probes from the earliest experiments on ultracold atoms. Similarly, thermal activation of collective modes for superfluid Helium droplets were investigated \cite{dst01} . The collective excitation frequencies of a trapped cloud reveal information about the equilibrium state and are in general sensitive to local compressibility of the system. For a short range interacting BEC, the collective oscillation frequencies are only weakly dependent on temperature except near the BEC critical temperature \cite{jme97}. However, the dipolar droplet state is stabilized by fluctuations and the equilibrium can be more sensitive to thermal effects.

The oscillation modes for a dipolar gas at zero temperature has been calculated in \cite{rbb06}. For the droplet state the lowest lying oscillation modes have been calculated in \cite{wsa16-2} by a variational approach, while the more general spectrum of excited modes of the droplet have been explored numerically by \cite{bwb16-2}.  The oscillation frequencies have been measured for Er droplets \cite{cbp16}, where the signature of the local LHY correction was clearly identified in the axial oscillation mode frequency. Similarly, collective oscillations in Dy droplets have been excited by tilting the dipole orientations \cite{fwb18}.

We explore the few lowest lying modes of the droplet through both approaches. First, we write a variational Lagrangian for a Gaussian wavefunction for which coupled oscillations of the cloud width in three dimensions give us the lowest three modes of the condensate. We compare the values found from this approach with numerical real time evolution of the condensate.
\begin{widetext}

We use a Gaussian ansatz for the wavefunction, 
\begin{align}
\Psi(x,y,z,t) &= \frac{\sqrt{N}}{\pi^{3/4}\left(w_{x}(t)w_{y}(t)w_{z}(t)\right)^{1/2}} \\
& \exp \left[-\frac{1}{2}\left(\frac{x^{2}}{w_{x}(t)^{2}}+\frac{y^{2}}{w_{y}(t)^{2}}+\frac{z^{2}}{w_{z}(t)^{2}}\right) + i\left(x^{2}\beta_{x}(t)+y^{2}\beta_{y}(t)+z^{2}\beta_{z}(t)\right) \right]
\end{align}
where the time dependent parameters $w_{x,y,z}$ give the cloud radii. The other variational parameters $\beta_{x,y,z}$ facilitate the mass current so that the cloud density can oscillate. Such a restricted form for the variational wavefunction can only capture the three collective oscillation frequencies of the condensate. However, these lowest modes have been the most accessible modes experimentally and variational Lagrangian method results have been in good agreement with measurements \cite{cbp16,wsa16-2}.

For elongated cigar shape condensates such as the dipolar droplets, one of these modes feature oscillations predominantly in the axial direction. This axial mode has the lowest frequency which has been measured in Er and Dy experiments. The other two modes consist of breathing and quadrupolar excitations perpendicular to the long axis and typically have frequencies an order of magnitude larger than the axial mode for the droplets.

The effect of the beyond mean field correction for the oscillation frequencies have first been reported in \cite{wsa16-2}. We now introduce the temperature dependence into the LHY correction. Our Lagrangian is:
\begin{align}
  \mathcal{L}=&\frac{i\hbar}{2}\left(\Psi(\mathbf{x},t)\frac{\partial \Psi^{*}(\mathbf{x},t)}{\partial t}-
  \Psi^{*}(\mathbf{x},t)\frac{\partial \Psi(\mathbf{x},t)}{\partial 
  t}\right)+\frac{\hbar^{2}}{2M}|\nabla\Psi(\mathbf{x},t)|^{2}\\ \nonumber
  &+U_{\text{tr}}(\mathbf{x})|\Psi(\mathbf{x},t)|^{2} \\\nonumber
  &+\frac{1}{2}\int d^{3} \mathbf{x'}|\Psi(\mathbf{x},t)|^{2}V_{\text{int}}(\mathbf{x}-\mathbf{x'})
  |\Psi(\mathbf{x'},t)|^{2} \\\nonumber
  &+\frac{2}{5}\gamma |\Psi(\mathbf{x},t)|^{5} \\\nonumber
  &+2\theta T^{2} |\Psi(\mathbf{x},t)|,
\end{align}
where $\gamma = 
\frac{32}{3}g\sqrt{\frac{a_{s}^{3}}{\pi}}\mathcal{Q}_{5}(\epsilon_{dd})$, and 
$\theta = 
\frac{32}{3}g\sqrt{\frac{a_{s}^{3}}{\pi}}\frac{k_{B}^{2}}{g^{2}}\mathcal{S}(\epsilon_{dd})$

The integrals in the Lagrangian can be carried out for the variational wavefunction to yield the Lagrangian in terms of the variational parameters as
\begin{align}
  L(w_{x},w_{y},w_{z},\beta_{x},\beta_{y},\beta_{z})  = &
   \frac{N\hbar}{2}\left(w_{x}^{2}\dot{\beta}_{x}+w_{y}^{2}\dot{\beta}_{y}+w_{z}^{2}\dot{\beta}_{z}\right) \\\nonumber
   &+ \frac{\hbar^{2}N}{4M} \left(\frac{1}{w_{x}^{2}}+\frac{1}{w_{y}^{2}}+
   \frac{1}{w_{z}^{2}}+4w_{x}^{2}\beta_{x}^{2}+4w_{y}^{2}\beta_{y}^{2}+4w_{z}^{2}\beta_{z}^{2}\right) \\\nonumber
   &+ \frac{1}{4}MN \left(\omega_{x}^{2}w_{x}^{2}+\omega_{y}^{2}w_{y}^{2}+\omega_{z}^{2}w_{z}^{2}\right) \\\nonumber
   &+ \frac{gN^{2}}{2}\frac{1}{2^{3/2}\pi^{3/2}w_{x}w_{y}w_{z}}\left(1-\epsilon_{dd}
   F\left(\frac{w_{x}}{w_{z}},\frac{w_{y}}{w_{z}}\right)\right) \\ \nonumber
   &+ \frac{4\sqrt{2}}{25\sqrt{5}\pi^{9/4}} 
  \frac{N^{5/2}\gamma}{\left(w_{x}w_{y}w_{z}\right)^{3/2}} \\ \nonumber
  &+ 4\sqrt{2}\pi^{3/4}\theta T^{2} N^{1/2}\left(w_{x}w_{y}w_{z}\right)^{1/2}.
\end{align}
where  \begin{equation}
    F(\kappa_{x},\kappa_{y})=1-3\kappa_{x}\kappa_{y}\int_{0}^{\pi}\frac{d\phi}{\pi}\left(\frac{\tanh^{-1}\left[\sqrt{1-\left(\kappa_{x}^{2}\cos^{2}\phi+\kappa_{y}^{2}\sin^{2}\phi\right)}\right]}{\left(1-\left(\kappa_{x}^{2}\cos^{2}\phi+\kappa_{y}^{2}\sin^{2}\phi\right)\right)^{3/2}}-\frac{1}{1-\left(\kappa_{x}^{2}\cos^{2}\phi+\kappa_{y}^{2}\sin^{2}\phi\right)}\right).
  \end{equation}

The Euler Lagrange equations for the radii give:
\begin{equation}
\hbar w_{\eta}\dot{\beta}_{\eta} +  \frac{2\hbar^{2}}{M}w_{\eta}\beta_{\eta}^{2} 
  + \frac{\partial G}{\partial w_{\eta}}=0,
\end{equation}
while the phase terms are simply related to the velocity of the radii as
\begin{equation}
  \beta_{\eta}=\frac{M}{2\hbar w_{\eta}}\dot{w}_{\eta},
\end{equation}
where,
\begin{align}
  G(w_{x},w_{y},w_{z}) =& \frac{\hbar^{2}}{4M} \left(\frac{1}{w_{x}^{2}}+\frac{1}{w_{y}^{2}}+
   \frac{1}{w_{z}^{2}}\right) + \frac{1}{4}M \left(\omega_{x}^{2}w_{x}^{2}+\omega_{y}^{2}w_{y}^{2}+\omega_{z}^{2}w_{z}^{2}\right)
  \\\nonumber & + \frac{gN}{2}\frac{1}{2^{3/2}\pi^{3/2}w_{x}w_{y}w_{z}}\left(1-\epsilon_{dd}
   F\left(\frac{w_{x}}{w_{z}},\frac{w_{y}}{w_{z}}\right)\right) \\ \nonumber &+
    \frac{4\sqrt{2}}{25\sqrt{5}\pi^{9/4}} 
  \frac{N^{3/2}\gamma}{\left(w_{x}w_{y}w_{z}\right)^{3/2}} +
  4\sqrt{2}\pi^{3/4}\theta T^{2} N^{-1/2}\left(w_{x}w_{y}w_{z}\right)^{1/2}.
\end{align}
Therefore,
\begin{equation}
  \frac{d^{2} w_{\eta}}{d t^{2}} = - \frac{2}{M} \frac{\partial}{\partial w_{\eta}} G(w_{x},w_{y},w_{z}).
\end{equation}
We non-dimensionalize the equations in terms of the effective trap frequency $\bar{\omega}=\left(\omega_{x}\omega_{y}\omega_{z}\right)^{1/3}$, and the harmonic oscillator length $\bar{l}=\sqrt{\frac{\hbar}{M\bar{\omega}}}$. 
With $\nu_{i}=w_{i}/\bar{l}$, $\tau=\bar{\omega}t$ and $E_{0} = \hbar \bar{\omega}$, the effective potential $\tilde{U}$ becomes
\begin{align}
  \tilde{U} =& \frac{1}{2} \left(\frac{1}{\nu_{x}^{2}}+\frac{1}{\nu_{y}^{2}}+
   \frac{1}{\nu_{z}^{2}}\right) + \frac{1}{2} \left(\frac{\omega_{x}^{2}}{\bar{\omega}^{2}}\nu_{x}^{2}+
   \frac{\omega_{y}^{2}}{\bar{\omega}^{2}}\nu_{y}^{2}+
   \frac{\omega_{z}^{2}}{\bar{\omega}^{2}}\nu_{z}^{2}\right)
  \\\nonumber& + \sqrt{\frac{2}{\pi}}\frac{Na_{s}}{\bar{l}}\frac{1}{\nu_{x}\nu_{y}\nu_{z}}\left(1-\epsilon_{dd}
   F\left(\frac{\nu_{x}}{\nu_{z}},\frac{\nu_{y}}{\nu_{z}}\right)\right) \\\nonumber&+
    \mathbf{c}_{1}\frac{N^{3/2}a_{s}^{5/2}}{\bar{l}^{5/2}}
  \frac{1}{\left(\nu_{x}\nu_{y}\nu_{z}\right)^{3/2}} 
  \mathcal{Q}_{5}(\epsilon_{dd})\\\nonumber
  &+
  \mathbf{c}_{2}\frac{a_{s}^{1/2}}{N^{1/2}\bar{l}^{1/2}}
  \left(\nu_{x}\nu_{y}\nu_{z}\right)^{1/2}
  \tilde{t}^{2}\mathcal{S}(\epsilon_{dd}),
\end{align}
\end{widetext}
where $\mathbf{c}_{1}=\frac{1024\sqrt{2}}{75\sqrt{5}\pi^{7/4}}\approx 1.1648$, 
and $\mathbf{c}_{2}=\frac{64\sqrt{2}}{3\pi^{3/4}}\approx 12.785$.
The equilibrium radii are found at the stationary point of the effective potential
\begin{equation}
  \left.\frac{\partial \tilde{U}}{\partial \nu_{\eta}} 
  \right|_{\nu_{\eta (0)}} = 0.
\end{equation}
Taylor expansion to second order near the equilibrium point defines the Hessian matrix 
\begin{equation}
  M = \begin{bmatrix}
               k_{x}&\lambda_{xy}&\lambda_{zx}\\
                \lambda_{xy}&k_{y}&\lambda_{yz}\\
                \lambda_{zx}&\lambda_{yz}&k_{z}
           \end{bmatrix},
\end{equation}
where $k_{\eta} = \left.\frac{\partial^{2}\tilde{U}}{\partial \nu_{\eta}^{2}}\right|_{(0)}$, 
$\lambda_{\alpha,\beta} = \left.\frac{\partial^{2} \tilde{U}}{\partial \nu_{\alpha} \partial 
\nu_{\beta}}\right|_{(0)}$. Collective frequencies are found via diagonalization of this matrix. We numerically calculate the derivatives near the equilibrium point and obtain the collective oscillation frequencies.
   
\begin{figure}
\centering

    \includegraphics[scale=0.55]{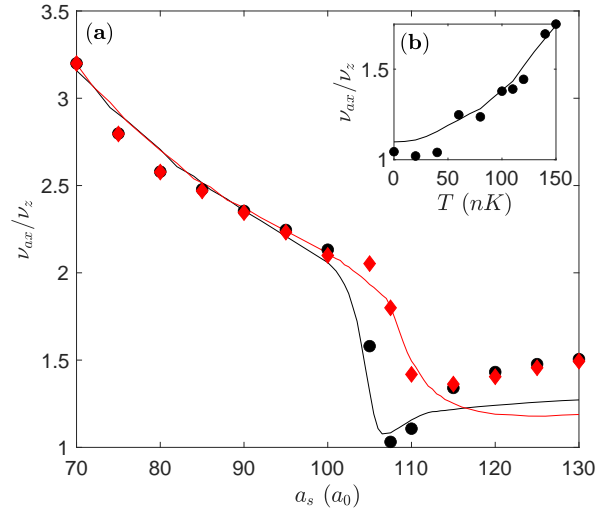}
    \caption{(a) Axial oscillation frequencies of a Dy BEC with $N=2000$ atoms in a pancake shaped trap with harmonic frequencies $\left\{\omega_\rho, \omega_z\right\}=2\pi\times\left\{70,70\right\}$ Hz as a function of the s-wave scattering length $a_s$. The black (lower) line and dots correspond to the temperature $T=0$, the red (upper) line and diamonds correspond to $T=150$nK. The lines are obtained analytically by assuming a Gaussian ansatz and the points are the results of numerical time evolution. Around the transition point to a high density droplet, the effect of temperature on the axial oscillations is significant and the frequencies are altered by around $75\%$. At a critical $a_s$ value of $108a_0$ (b) shows the change of the axial oscillation frequency as a function of the temperature. Both figures show the frequencies in units of the axial trap frequency.}
    
    \label{fig:DyOscillations}
\end{figure}

As a check on our variational results we also calculate the oscillation frequencies numerically through real time evolution with the split step method outlined in the previous section. We first calculate the oscillation frequencies for the Dy experimental parameters for which there is an abrupt transition. The oscillation frequencies increase upon transition to the droplet state, as can be seen in Fig.\ref{fig:DyOscillations}. For $20000$ Dy atoms in a spherical trap with $f=75$Hz the transition takes place near $a_s\simeq 105 a_0$ and is signalled by an abrupt change in the axial oscillation frequency.  Repeating the same calculation for $T=150$nK we observe that the  transition in the frequency becomes smoother as the transition point moves to higher $a_s$. Thus the most dramatic effect of temperature on the oscillation frequencies happens near the transition. The change in the frequency of the axial mode as a function of temperature is displayed in Fig. \ref{fig:DyOscillations}, where we observe close to 70\% change near the critical $a_s=105 a_0$.  Both variational and numerical results give parallel physical pictures indicating that the increase in the oscillation frequency with temperature is greatest near the transition to the droplet state.

\begin{figure}

\centering
    \includegraphics[scale=0.55]{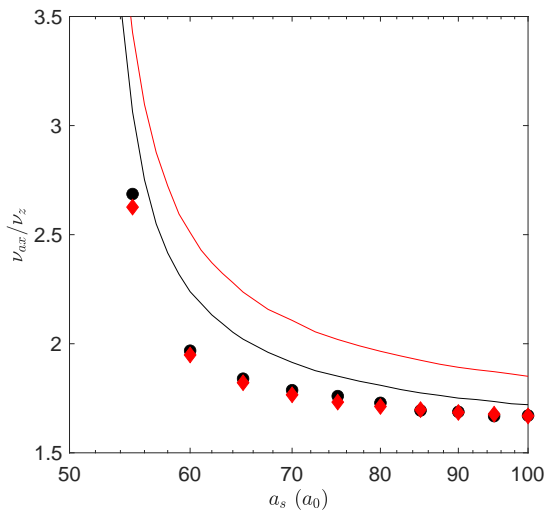}
    \caption{Axial oscillation frequencies of an Er BEC with $N=20000$ atoms in a cigar shaped trap with harmonic frequencies $\left\{\omega_\rho, \omega_z\right\}=2\pi\times\left\{178,17\right\}$ Hz as a function of the s-wave scattering length $a_s$. The oscillation frequency is in the units of axial trap frequency. The black (lower) line and dots correspond to the temperature $T=0$, the red (upper) line and diamonds correspond to $T=150$nK. The lines are obtained analytically by assuming a Gaussian ansatz and the dots are the results of numerical time evolution. For the cigar shaped droplets the low energy excitations are not altered significantly by the finite temperature effects.}
    \label{fig:ErOscillations}
\end{figure}

In the Er experiments the cigar shaped condensate has a smooth transition between the low density and droplet states. This smooth transition causes the cloud shape to be mostly immune to changes in temperature as explored in the previous section. Both the time dependent variational approach and the numerical real time evolution predict a sharp rise in the axial mode frequency upon droplet formation, which agree with the experimental observations in Ref.\cite{cbp16}. The variational approach predicts an increase in the oscillation frequencies with temperature as can be seen in Fig.\ref{fig:ErOscillations}. However numerical solutions of the same equation show very little change from the zero temperature frequencies up to 150nK. As the droplet state is extremely elongated the variational approach may be overstating the importance of the compressibility of the center of the cloud by disregarding higher modes along the axial direction. Thus, we expect no significant low-temperature effect on the axial oscillation frequency close to smooth crossover to the droplet state.

\section{Relevance for recent experiments} 
\label{sec:experiments}

In this paper we provide further support for the claim that dipolar droplets are particularly susceptible to thermal fluctuations at low temperatures. Numerical calculations presented here support the conclusions of the previous variational calculation \cite{aok19}. Furthermore, we show that the oscillation frequencies of collective modes are also sensitive to the temperature near the droplet transition. This should be expected as the collective modes are good probes of compressibility and the primary effect of temperature is to change the droplet's local chemical potential.   

Standard experimental methods for thermometry are not applicable to the current droplet experiments. The small size of the droplets impede the in-situ density measurement of the thermal component. Long range dipolar interactions affect the free expansion of the condensate, even when the droplet is not self-trapped \cite{bws19}. The temperature data given in the literature is based on the thermal component before droplet formation for both the Dy \cite{ksw16} and the Er systems \cite{cbp16}. In the absence of reliable temperature measurement it is not possible to make a direct comparison of our results with the experimental data, either for Dy or Er experiments. Our results show that a change in the oscillation frequencies can occur even at temperatures much lower than the BEC transition temperature, and this may explain the uncertainty in the range of frequencies observed in  \cite{fwb18,cbp16}. A change in the oscillation frequency for systems intentionally prepared at a higher temperature, or heated after droplet formation by an external probe would provide convincing evidence for the role of thermal fluctuations on dipolar droplets.

It is important to list the limitations of the approach used here.  Our theory does not take the interactions between excited (above the condensate) particles into account, which limits it's applicability to systems where the number of thermal atoms is close to the number of atoms in the condensate. Fluctuations are treated in the local density approximation, hence the independent dynamics of the thermal cloud is not considered. Furthermore, the expression for the local chemical potential suffers from an ad-hoc cutoff. These limitations make it hard to propose the collective oscillation frequencies as a quantitative measure of temperature. However, we believe that  the dependence of the oscillation frequencies on temperature should be observable in the current experimental regime.

This project is supported by T\"{u}rkiye Bilimsel ve Teknolojik Ara\c{s}t{\i}rma Kurumu
(T\"{U}B\.{I}TAK) Grant No. 116F215.

\bibliography{ThermalDropletsVersion2}
\end{document}